\documentclass[
    ,final            % use final for the camera ready runs
  ,numberedheadings % uncomment this option for numbered sections
  ]
  {aipproc}

\layoutstyle{8x11single}

\usepackage{latexsym,amsfonts,amsmath,amssymb,url,dsfont}
\usepackage{graphicx}
\usepackage{boxedminipage}
\usepackage{xspace}
\usepackage{color}             % Requis par xfig
\usepackage{enumerate}
\usepackage{subfig}
\usepackage{epsfig}
\usepackage{multirow}
\usepackage{wrapfig}

\usepackage{aas_macros}

\newcommand{\etc}{\emph{etc.}\xspace}
\newcommand{\eg}{\emph{e.g.,}\xspace}
\newcommand{\ie}{\emph{i.e.,}\xspace}

\newcommand{\diet}{\textsc{Diet}\xspace}
\newcommand{\dietwebboard}{\textsc{DietWebboard}\xspace}
\newcommand{\dagda}{\textsc{Dagda}\xspace}
\newcommand{\madag}{\textsc{MA$_{DAG}$}\xspace}

\newcommand{\corba}{CORBA\xspace}
\newcommand{\sed}{\textsc{SeD}\xspace}
\newcommand{\seds}{\textsc{SeD}s\xspace}

\newcommand{\ramses}{\textsc{Ramses}\xspace}
\newcommand{\gadget}{\textsc{Gadget}\xspace}
\newcommand{\galics}{\textsc{Galics}\xspace}
\newcommand{\galaxymaker}{\textsc{GalaxyMaker}\xspace}
\newcommand{\galaxymakers}{\textsc{GalaxyMakers}\xspace}
\newcommand{\momaf}{\textsc{MoMaF}\xspace}

\begin{document}

\title{Cosmological Simulations on a Grid of Computers}
%\title{Parameter Sweep Cosmological Simulations}

\classification{98.80.Es, 97.75.Pq}
\keywords      {Cosmology, N-body simulations, Parallel computing, Local universe.}
{\bf  Accepted and Published in AIP Conference Proceedings 1241, pages 816-825}

\author{Benjamin Depardon}{
  address={University of Lyon; ENS-Lyon/INRIA/CNRS/UCBL; LIP
Laboratory. France.}
}

\author{Eddy Caron}{
  address={University of Lyon; ENS-Lyon/INRIA/CNRS/UCBL; LIP
Laboratory. France.}
}

\author{Fr{\'e}d{\'e}ric Desprez}{
  address={University of Lyon; ENS-Lyon/INRIA/CNRS/UCBL; LIP
Laboratory. France.}
}

\author{J{\'e}r{\'e}my Blaizot}{
  address={University of Lyon; UCB Lyon 1/CNRS/INSU; CRAL. France.}
}

\author{H{\'e}l{\`e}ne Courtois}{
  address={University of Lyon; UCB Lyon 1/CNRS/IN2P3/INSU; IPN Lyon. France.}
}

\begin{abstract}

  The work presented in this paper aims at restricting the input
  parameter values of the semi-analytical model used in \galics and
  \momaf, so as to derive which parameters influence the most the
  results, \eg star formation, feedback and halo recycling
  efficiencies, \etc Our approach is to proceed empirically: we run
  lots of simulations and derive the correct ranges of values. The
  computation time needed is so large, that we need to run on a grid
  of computers. Hence, we model \galics and \momaf execution time and
  output files size, and run the simulation using a grid middleware:
  \diet. All the complexity of accessing resources, scheduling
  simulations and managing data is harnessed by \diet and hidden
  behind a web portal accessible to the users.

\end{abstract}

\maketitle

%%%%%%%%%%%%%%%%%%%%%%
\section{Introduction}
\label{sec:introduction}

Cosmological simulations in this context are used to simulate the
evolution of dark matter through cosmic time in various universes. A
classical simulation begins with an N-body computation using for
example \ramses~\cite{2002AA_385_337T} or
\gadget~\cite{2005MNRAS_364_1105S}. The output of the simulation is
then post-processed using semi-analytical models such as
\galics~\cite{2003MNRAS_343_75H} (\textbf{GAL}axies \textbf{I}n
\textbf{C}osmological \textbf{S}imulations), then mock catalogs of
observed galaxies are produced using \momaf~\cite{2005MNRAS_360_159B}
(\textbf{Mo}ck \textbf{Ma}p \textbf{F}acility). Those models use as
input a set of parameters, which influence the results, as for
example: the galaxy luminosity function, the number of galaxies per
observed cones, and more globally the history of star formation rate
in the galaxy evolution process. Those parameters have a large range
of values, the experiment here aims at reducing for each parameter the
intervals of values to a subset of ranges within which the output
simulations would be coherent with observed galaxies distributions.

In order to stress how realistic the post-processing results are, a
first important step is to identify the \galics and \momaf parameters
which have the largest impact on the astrophysical results, such as
star formation efficiency, feedback efficiency, halo recycling
efficiency. As the parameters have a large range of possible values,
exploring ``all'' combinations is really time consuming and requires
lots of computing power. To harness the difficulty of running these
analysis, one need firstly to have access to many computing resources,
and secondly an efficient way to access those latter. Hence, we
propose a client/server implementation for running post-processings on
a distributed platform composed of heterogeneous machines: a
\emph{Grid}. A transparent access to these machines is provided by a
\emph{grid middleware}: \diet (\textbf{D}istribued
\textbf{I}nteractive \textbf{E}ngineering \textbf{T}oolbox). \diet
handles in one common and effective way the deployment of the
computations on a grid of heterogeneous and distributed computers, the
management of the different components of the post-processing, and
also provides monitoring, communications and computation scheduling,
and data and workflows management.

Running efficiently the post-processings on a set of distributed and
heterogeneous machines requires estimations on both the execution time
and the amount of data to be transfered for each applications. Hence,
we benchmarked \galics and \momaf and derived their execution time and
output file size. Those models are used within \diet to select which
computer should run the post-processing.

We first present in Section~\ref{sec:cosmo} the cosmological
simulation post-processing workflow, then we give the execution time
and output file size models in Section~\ref{sec:model}. In
Section~\ref{sec:ArchDIET} we give an overview of the \diet
middleware: its architecture, and the various features used within
this project. Finally, and before concluding the paper, we present in
Section~\ref{sec:implem} the client/server implementation that allows
the transparent execution of the post-processing workflow.

%%%%%%%%%%%%%%%%%%%%%%%%%%%%%%%%%
\section{Cosmological Simulations Post-Processing}
\label{sec:cosmo}

%% Cosmological Simulations are CPU and  disk intensive
%% applications. They are sequential and parallel. The workflow can be
%% distributed across several machines.

%% Running Cosmological Simulations on a grid implies the use of
%% distributed platforms, which means heterogeneity between the different
%% processors.

%% The goal of such an experiment is to stress \diet and the GRID'5000
%% using real scientific applications on a long period of time.

Post processing cosmological simulations is done in two steps: we
first need to run \galaxymaker, and then \momaf. As there isn't only a
single input file, and as the intermediary files are numerous and used
several times, \ie the output of \galaxymaker can be processed by
possibly many \momaf instances, we represent the whole execution by a
workflow, \ie a graph depicting dependencies between the different
tasks.

\begin{figure}[h]
  \centering
  \includegraphics[width=.5\textwidth]{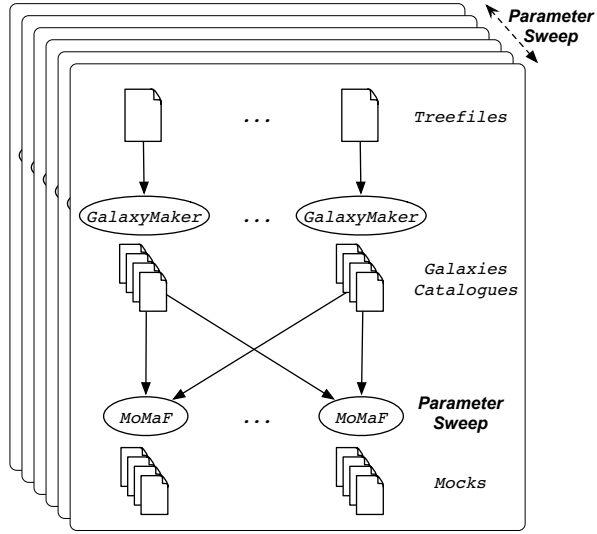}
  \caption{Workflows' pattern. A workflow can be executed many times on
    several parameter sets.
  \label{fig:cosmo_wf}}
\end{figure}

Figure~\ref{fig:cosmo_wf} presents the workflows' pattern. The input of
each \galaxymaker is a treefile (\ie a file containing the merger
trees of the halos of dark matter)  and a set of astrophysical
parameters. In order to parallelize the hierarchical galaxy formation
computation, this file can be divided into several smaller files, fed
to numerous \galaxymaker instances. Then, all \galaxymaker's outputs
(\ie list of galaxies) have to be post-processed by \momaf in
order to produce mock catalogs of observed galaxies.

In order to explore the parameter space, and provide different views
for the virtual observations, both \galaxymaker and \momaf need to be
run with different sets of parameters. We have the following variable parameters:
\paragraph{\galaxymaker variable parameters}
  \begin{itemize}
  \item star formation efficiency ($alphapar$)
  \item feedback efficiency ($epsilon$)
  \item halo recycling efficiency ($upsilon$)
  \item SN feedback model ($silk feedback$: $true$ = Silk 01, $false$ = SP99)
  \end{itemize}
  
\paragraph{\momaf variable parameters}
  \begin{itemize}
  \item the opening angle in the right ascension and declination
    directions (respectively $ra\_size$: from 0 to 360 degrees, and
    $dec\_size$: from 0 to 179.9 degrees).
  \item the minimum and maximum comoving distance (comoving distance
    is the distance between two points measured along a path defined
    at the present cosmological time) to observer, \ie all objects
    outside this interval is excluded (respectively $min\_depth$ and
    $max\_depth$)
  \end{itemize}

A given set of parameters for \galaxymaker produces one workflow: each
treefiles are processed by an instance of \galaxymaker, each instance
having the same set of parameters. However, different parameter sets
can be fed to \momaf within a same workflow: all \galaxymaker's
outputs are fed to different instances of \momaf which will produce
different results. Hence, when using different parameter sets for both
\galaxymaker and \momaf, we obtain what is depicted on
Figure~\ref{fig:cosmo_wf}, we have parameter sweep at two levels:
within each workflow, and to generate several instances of the
workflows.

\section{Modeling \galaxymaker and \momaf}
\label{sec:model}

Having as much knowledge as possible on an application always helps
running it efficiently on a distributed platform. As said previously,
these applications are both computing and data intensive, and as the
goal of this work is to run cosmological simulations in a parameter
sweep manner, we studied the impact of each parameter on the execution
time, and on the output files' size for both \galaxymaker and
\momaf. Benchmarks were conducted on the Grid'5000 experimental
platform~\cite{Bolze:2006bx}, on one of the Lyon cluster: each node
has an AMD Opteron 250 CPU at 2.4GHz, with 1MB of cache and 2GB of
memory.

\subsection{\galaxymaker}

We ran \galaxymaker on input files containing the tree files from a
$512^3$ particles $100 Mpc.h^{-1}$ simulation. The variable parameters
were: the input file size, alphapar, epsilon, upsilon and silk
feedback.

\begin{figure}[ht]
  \centering
  \includegraphics[width=.62\textwidth]{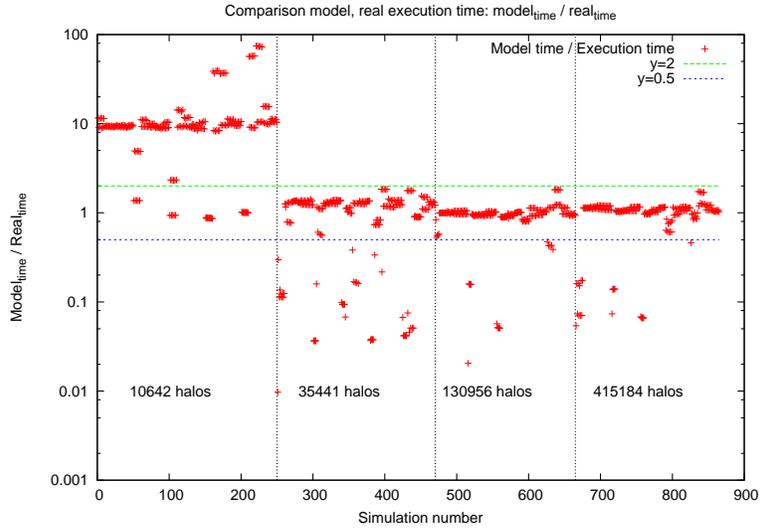}
  \caption{\galaxymaker model execution time error.
  \label{fig:galaxymaker_time}}
\end{figure}

Equation~\eqref{eq:galaxymaker_time} presents the execution time model
for \galaxymaker. Figure~\ref{fig:galaxymaker_time} presents the ratio
between the execution time given by the model, and the real execution
time. As can be seen, for small input files, the model overestimates
the execution time, this is often the case when modeling application
behavior on small inputs: there is less swapping and caching problems.

\begin{equation}
  \label{eq:galaxymaker_time}
  T_{\galaxymaker} = a_t \times nb_{halos} + b_t + c_t \times \left(\frac{alphapar}{epsilon}\right)^{d_t}
\end{equation}
Where $T$ is the execution time in seconds,
$nb_{halos}$ is the number of halos found in the input file (\ie this
is related to its size), $a_t$ and $b_t$ are constants, and $c_t$ and
$d_t$ are linear functions of $nb_{halos}$. We found the following
values:
\begin{itemize}
\item $a_t = 0.00563537$
\item $b_t = - 28.7845$
\item if silk feedback is $false$
  \begin{itemize}
  \item $c_t = 0.00946453 \times nb_{halos} + 71.6585$
  \item $d_t = 3.64906 \cdot 10^{-7} \times nb_{halos} + 1.44597$
  \end{itemize}
\item if silk feedback is $true$
  \begin{itemize}
  \item $c_t = 0.00441855 \times nb_{halos} - 40.6202$
  \item $d_t = -1.83813 \cdot 10^{-7} \times nb_{halos} + 1.80095$
  \end{itemize}
\end{itemize}
As can be seen, upsilon does not appear in the model, as it influences
only slightly the execution time.

\begin{figure}[ht]
  \centering
  \includegraphics[width=.62\textwidth]{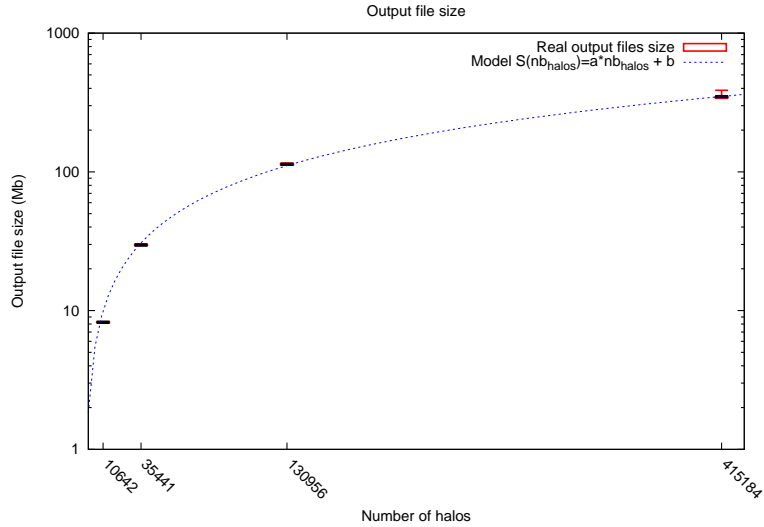}
  \caption{Galaxymaker output files size.
  \label{fig:galaxymaker_size}}
\end{figure}

The output files size was easier to model: it depends only on the
input file size, \ie the number of
halos. Equation~\ref{eq:galaxymaker_size} presents the model for
output file size, and Figure~\ref{fig:galaxymaker_size} the comparison
between the model and the real output files' size. The output files'
size is really stable, and thus the model perfectly matches the real
output size.
\begin{equation}
  \label{eq:galaxymaker_size}
  S_{\galaxymaker} = a_s \times nb_{halos} + b_s
\end{equation}
Where $S$ is the output files size in Mb, $a_s = 8.39317 \cdot
10^{-4}$ and $b = 9.99961 \cdot 10^{-1}$.

\subsection{\momaf}

\begin{figure}[ht]
  \centering
  \includegraphics[width=.62\textwidth]{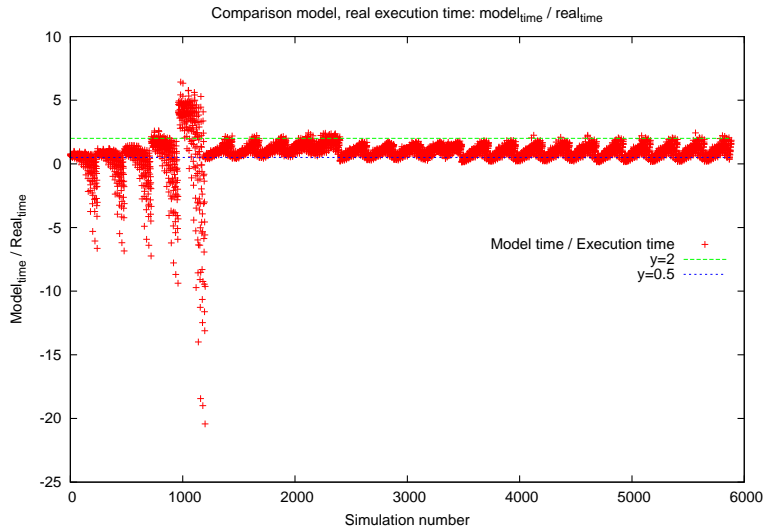}
  \caption{\momaf model execution time error.
  \label{fig:momaf_time}}
\end{figure}

Both the execution time and the output files size of this application
are less stable than \galaxymaker's. We weren't able to derive a model
that would fit for any input file, and any parameter. Hence, we only
present here the model for a given input file, and do not
give the values of the various parameters as one would need to find
them for each input file. Equation~\ref{eq:momaf_time} presents the
model, and Figure~\ref{fig:momaf_time} presents the ratio between the
model execution time, and the real execution time. Those were obtained
by varying the following parameters: opening angles in the declination
and ascension directions ($dec\_size$ and $ra\_size$), and minimum and maximum comoving distance to
observer ($min\_depth$ and $max\_depth$).

\begin{equation}
  \label{eq:momaf_time}
  T_{\momaf} = a_t \times \left(min\_depth - max\_depth\right)^{b_t} + c_t
\end{equation}
With:
\begin{itemize}
\item $a_t = \alpha_a \times ra\_size^{\beta_a} \times
  dec\_size^{\gamma_a} + \delta_a$
\item $b_t = \alpha_b \times \log\left(ra\_size\right) + \beta_b \times
  \log\left(dec\_size\right) + \gamma_b \times ra\_size + \delta_b
  \times dec\_size$
\item $c_t = \alpha_c \times ra\_size^{\beta_c} \times dec\_size^{\gamma_c} + \delta_c$
\end{itemize}
$\alpha_{a,b,c}$, $\beta_{a,b,c}$, $\gamma_{a,b,c}$, and
$\delta_{a,b,c}$ are constants.

\vspace{.5cm}
Exploring the whole parameter space, requires lots of computing power,
and data storage. Those can be provided by a grid of computers, which
has the advantage of having thousands of interconnected computers, but
has the drawback of being highly heterogeneous and distributed on a
large scale. Hence, accessing those machines is not as easy as
accessing machines on a cluster. Moreover, managing thousands of
workflows on a distributed environment can be really complex if merely
using scripts. Thus, we need a mean of running jobs transparently on a
grid, \ie a layer between the hardware and the software that hides the
complexity of the platform: we need a \emph{grid middleware}.

%%%%%%%%%%%%%%%%%%%%%%%%%%%%%%%%%
\section{Distributed Interactive Engineer Toolbox}
\label{sec:ArchDIET}

We now introduce \diet, a scalable distributed middleware for
accessing transparently and efficiently heterogeneous and highly
distributed machines.

\subsection{The \diet architecture}
 
The \diet component architecture is structured hierarchically for
improved scalability. Such an architecture is flexible and can be
adapted to diverse environments, including arbitrary heterogeneous
computing platforms.  The \diet
toolkit~\cite{Caron:2006gf,Amar:2008xy} (\textbf{D}istributed
\textbf{I}nteractive \textbf{E}ngineering \textbf{T}oolbox) is
implemented in \corba and thus benefits from the many standardized,
stable services provided by freely-available and high performance
\corba implementations. \corba systems provide a remote method
invocation facility with a high level of transparency.  This
transparency should not substantially affect the performance, as the
communication layers in most \corba\ implementations is highly
optimized~\cite{Denis:2001jt}.  These factors motivate our decision to
use \corba as the communication and remote invocation fabric in \diet.

\begin{figure}[h]
  \centering
  \includegraphics[width=.6\textwidth]{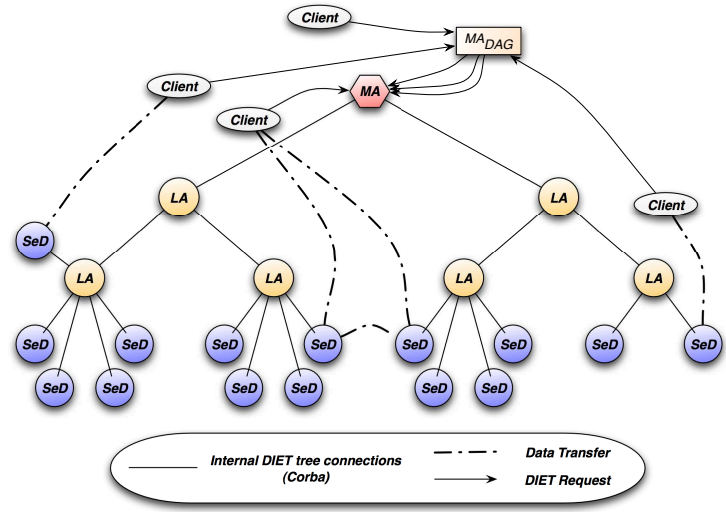}
  \caption{\diet hierarchical organization. Plug-in scheduler are
    available in each MA and LA. Clients can submit requests directly
    to the MA or submit workflows to the \madag.\label{fig:diet-overview}}
\end{figure}

The \diet framework comprises several components.  A \textbf{Client}
is an application that uses the \diet infrastructure to solve problems
using a remote procedure call (RPC) approach.  Clients access \diet
via various interfaces: web portals  programmatically using published
C or C++ APIs. A \textbf{\sed}, or server daemon, acts as the service
provider, exporting functionality via a standardized computational
service interface; a single \sed can offer any number of computational
services.  A \sed can also serve as the interface and execution
mechanism for either a stand-alone interactive machine or a parallel
supercomputer, by interfacing with its batch scheduling facility. The
third component of the \diet architecture, \textbf{agents}, facilitate
the service location and invocation interactions of clients and \seds.
Collectively, a hierarchy of agents provides higher-level services
such as scheduling and data management. These services are made
scalable by distributing them across a hierarchy of agents composed of
a single \textbf{Master Agent (MA)} and several \textbf{Local Agents
  (LA)}. Figure~\ref{fig:diet-overview} shows an example of a \diet
hierarchy.

The \textbf{Master Agent} of a \diet hierarchy serves as the
distinguished entry point from which the services contained within the
hierarchy can be logically accessed.  Clients identify the \diet
hierarchy using a standard \corba naming service.  Clients submit
requests~-- composed of the name of the specific computational service
they require and the necessary arguments for that service~-- to the
MA. The MA then forwards the request to its children, who subsequently
forward the request to their children, such that the request is
eventually received by all \seds in the hierarchy. \seds then evaluate
their own capacity to perform the requested service; capacity can be
measured in a variety of ways including an application-specific
performance prediction, general server load, or local availability of
datasets specifically needed by the application. \seds forward this
capacity information back up the agent hierarchy.  Based on the
capacities of individual \seds to service the request at hand, agents
at each level of the hierarchy reduce the set of server responses to a
manageable list of server choices with the greatest potential. The
server choice can be made specific to any kind of application using
plug-in schedulers at each level of the hierarchy.

\subsection{Data management}
\label{sec:dagda}

Data management is handled by \dagda~\cite{AHEMA2008} (\textbf{D}ata
\textbf{A}rrangement for \textbf{G}rid and \textbf{D}istributed
\textbf{A}pplication), it allows data explicit or implicit
replications and advanced data management on the grid such as data
backup and restoration, persistency, data replacement algorithm. A
\dagda component is attached to each \diet element and follows the
same hierarchical distribution. However, whereas \diet elements can
only communicate following the hierarchy order (those communications
appear when searching a service, and responding to a request), \dagda
components will use the tree to find data, but once the data is found,
direct communications will be made between the owner of the data and
the one which requested it. The \dagda component associates an ID to
each stored data, manages the transfers by choosing the ``best'' data
source according to statistics about the previous transfers time and
performs data research among the hierarchy. Just like \diet, \dagda
uses the \corba interface for inter-nodes communications.

\subsection{Workflows}
\label{sec:workflows}

A large number of scientific applications are represented by graphs of
tasks which are connected based on their control and data
dependencies. The workflow paradigm on grids is well adapted for
representing such applications and the development of several workflow
engines~\cite{gridAnt04,taverna04,pegasus05,DAGMan} illustrate
significant and growing interest in workflow management. Several
techniques have been established in the grid community for defining
workflows. The most commonly used model is the graph and especially
the Directed Acyclic Graph (\textsc{Dag}).

\diet introduces a new kind of agent: the \madag. This agent is
connected to the MA as can be seen
Figure~\ref{fig:diet-overview}. Instead of submitting requests
directly to the MA, a client can submit a workflow to the \madag, \ie
an XML file containing the whole workflow description. The \madag will
then take care of the workflow execution, and schedule it along with
all the other workflows present in the system, hence the system can
manage multiple workflows concurrently. Thus, the client only needs to
describe the workflow in an XML format, then feed in the input data,
and finally retrieve the output data when the workflow has finished
its execution.

\subsection{Web Portal}
\label{sec:webboard}

In order to ease job submissions, a web portal has been
developed\footnote{Portal for cosmological simulations submission:
  \url{http://graal.ens-lyon.fr:5544/Cosmo/}\\
Description of the available features: \url{http://graal.ens-lyon.fr/DIET/dietwebboard.html}}. It
hides the complexity of writing workflow XML files. The user describes
astrophysical parameters, and provides the filters list file, the
observation cones description file, and a tarball containing one or
many treefiles, then clicks on submit, the system takes care of the
communications with \diet: it creates the corresponding workflow
descriptions, submit them to \diet, retrieve the final results and
make them available via a webpage. In order to explore the parameter
space, the submission webpage allows for astrophysical and cones
parameters to define three values: the minimum and maximum values, and
the increase step that needs to be applied between these two
values. Thus, with the help of a single webpage, the user is able to
submit lots of workflows.

In case a submission fails, the job is submitted again after a fixed
period, this until the job finishes properly, or is canceled by the
user. Once finished, a tarball file containing all result files can be
downloaded. Depending on the option chosen at the submission stage,
this file contains output files of both \galaxymaker and \momaf, or
just \momaf.

Access to this system is protected by a login/password authentication,
so as to restrict access to applications and produced data.  This
system also provides independent parameter sweep jobs submissions for
both applications.

\section{Framework}
\label{sec:implem}

\subsection{Client}
\label{sec:client}

The main idea is to provide a transparent access to computing
services. End users shouldn't have to write a single line of code, or
XML to be able to use the submission system. Thus, the \diet client
only needs to parameters: a tarball file containing all treefiles to
process, and a file containing the parameters for both \galaxymaker
and \momaf. This parameter file uses the same syntax than the ones
used with \galaxymaker and \momaf, hence one only has to concatenate
\galaxymaker parameter file with \momaf parameter file and remove any
double. The client in itself does not allow parameter sweep for
\galaxymaker, one has to call the client once for each \galaxymaker
parameter set. However, parameter sweep is provided for \momaf. This
behavior reflects the workflow description depicted on
Figure~\ref{fig:cosmo_wf}.

The client automatically creates the workflow description XML: for
each treefile a \galaxymaker service is added, and for each \momaf
parameter set, a \momaf service is added. The output of all
\galaxymakers are fed into each single instance of \momaf.

The workflow is then submitted to \diet. Data transfers and services
execution are automatically handled. Once the execution has ended, all
output data is retrieved at the client level: for each \galaxymaker
and \momaf output data is stored in an independent directory. If the
client is called from the \dietwebboard, all produced data is  compressed
and send back to \dietwebboard's storage disk and made available on
the webpage.

Cosmological simulations generate a lot of temporary files locally at
the client level, as well as on every server that executed a
\galaxymaker or \momaf service. Thus, once everything as been
executed, and useful data retrieved, the client calls cleaning
services which deletes now obsolete data on each server, this
mechanism is presented in more details in the following section.

\subsection{Server}

In this section we describe the mechanisms used by the servers in
order to manage data, and in order to efficiently access resources.

\paragraph{Data management}

These cosmological applications require as input large amounts of
data, the outcome of the processing is also very large. Depending on
the input parameters, as well as the number of filters applied in the
simulation, the number of files can vary. A problem arises: how to
transfer efficiently data between \galaxymaker, \momaf and the client.
Part of the data produced at one step need to be transmitted
efficiently to the next step, while some of them need to be sent back
to the client. Creating an archive by compressing produced files could
be a solution to deal with so many files, however, this would lead to
increasing disk usage (at least temporary, as it would require twice
the disk space used by the files at a given time), as well as
computation time (compressing or even storing files in an archive can
be costly). The option we chose is to have all produced data added
into a particular \diet data type: a container. A container can
contain any kind of \diet data (be it a file, an integer, a
vector, \dots), and we can add as many element as needed: the size is
dynamically managed. This structure allows to have an undetermined
number of files handled in a single data type, and thus will be
efficiently managed by \diet when data movement is required.

%\begin{wrapfigure}{l}{60mm}
\begin{figure}[h!t]
  \centering
  \includegraphics[width=.5\textwidth]{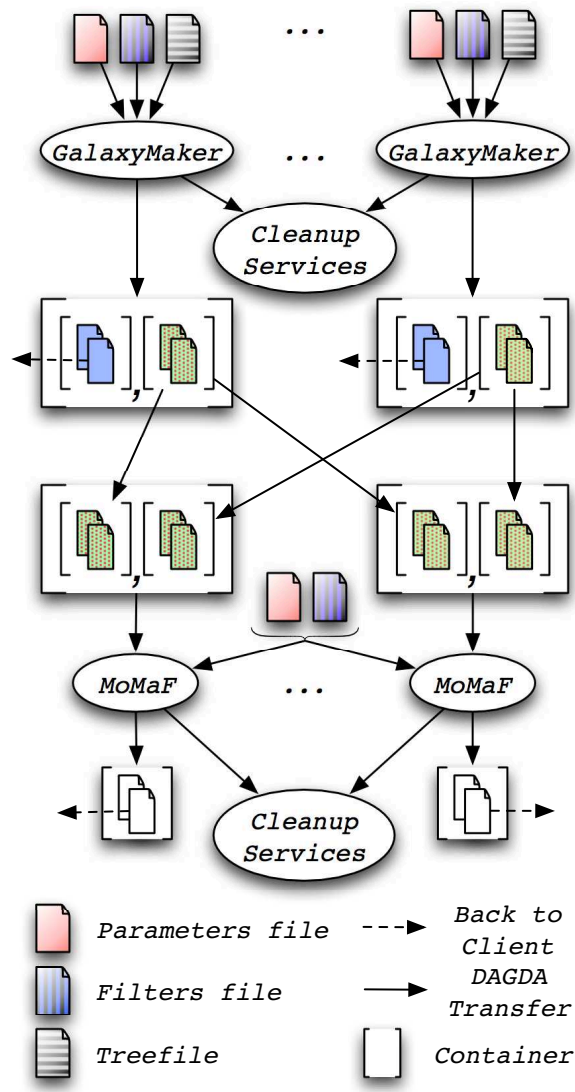}
  \caption{Data management.\label{fig:data_management}}
\end{figure}
%\end{wrapfigure}

Data is managed by \dagda so as to benefit from data persistency and
replication. Produced data is kept locally at \sed level, and is
downloaded on demand by the \sed executing the following service, and/or
the client.

As said previously, cosmological simulations produce a lot of data
that has to be kept during the whole workflow execution. Indeed, each
\galaxymaker output has to be post-processed by possibly many \momaf
instances.  A server only has local knowledge on the platform: it does
not know anything about the whole workflow, nor about the other
services' needs in terms of data: once a service has executed, it has
no means of knowing whether a data it has produced has already been
downloaded or not, and whether it will be needed in the future, hence
it has to keep. As a server cannot know by no means if a workflow has
ended, data would possibly stay indefinitely on the server, which is
of course impossible due to disk space constraints. One way of dealing
with this problem would be to let \dagda handle data removal when the
disk is full, according to a specified replacement algorithm
(currently three such algorithms are available: Least Recently Used,
Least Frequently Used and First In First Out). However this does not
solve our problem as \dagda could remove data still in use. To cope
with this problem, a new feature has been added to \diet. Usually a
service is defined by the \sed programmer in a static manner, this
service is instantiated when the \sed is run, and ends when the \sed
is killed. This does not reflect the dynamic behavior of our workflow
executions. Thus we added the possibility to add and remove services
dynamically at runtime. Thus, whenever a service is called, the \sed
keeps track of all created files and directories. Just before the
service ends, a cleanup service is spawned: its service name is unique
for the whole \diet platform (the service name is generated using a
Universally Unique Identifier, UUID, giving us reasonable confidence
that the service name is unique), and it requires no parameters, as
all required information is kept by the \sed. When this service is
called (\ie when the \madag detects that no services need the data
anymore), it deletes the corresponding previously registered files and
directories. Furthermore, in order to prevent many calls to this
cleaning service, the service automatically removes itself from the
list of available services. Hence, the machine is left clean, with no
temporary files remaining on the
disk. Figure~\ref{fig:data_management} depicts the data management
process when executing the whole workflow.

\paragraph{Accessing resources}

In order to be able to execute a service, a \sed need to access
computing resources. It has in fact two ways to do so: either the
service is run locally on the machine the \sed is deployed, or the
\sed can interact with a Batch system in order to submit a job to a
cluster or a grid. In the former case the service only has access to
the machine on which the \sed runs, but in the latter case, a service
can be run in parallel on many machines at the same. Note however that
if one has control on a cluster, nothing prevents her from deploying a
\sed per node on the cluster, and thus \diet won't have to deal with
the Batch system.

When interacting with the Batch system, the \sed needs to be deployed
where the Batch system runs, \eg the gateway, in order to be able to
communicate with it. In this case, the services aren't executed on the
gateway, but a node is requested to the batch system, \ie a script is
submitted to the system. This method complicates data management, as
once the script is submitted it cannot directly interact with \diet,
hence data has to be retrieved first by the \sed on the gateway in a
local directory (or a shared directory if available), then, when the
reservation is available, \ie the script starts running, required data
is sent to the corresponding node, where the application is
executed. Finally, when the service ends, the script sends all
produced data back to the gateway. The \sed detects the script
termination, adds produced data to \dagda so that the client and/or
the next step can retrieve it.

In order to be able to make reservations on the batch system, we need
an estimation of the job running time (\ie whenever a reservation is
made on a batch system, one has to specify its duration). As presented
in Section~\ref{sec:model}, an execution time model was obtained via
benchmarking \galaxymaker and \momaf with various parameter sets. Even
if these models aren't totally accurate (there may be an error factor
of 2, or even greater for small data sets), they give us a rough
estimation that can be used for submitting jobs to a batch system, as
well as for scheduling purposes between workflows: the \madag uses the
HEFT (Heterogeneous Earliest Time First)
heuristic~\cite{Topcuouglu:2002gd} to schedule the different
tasks. Hence, whenever a request is submitted, each \sed estimates the
execution time it would take to execute the request based on the model
and the previous executions. An history is kept locally at each \sed
in order to dynamically correct the model parameters: at the end of
each correct execution, the \sed updates a frame based history (it
keeps the bias between the model and the real execution time for the
last $n$ executions), and uses this history to predict the next
execution time. This techniques copes with two problems: errors induced
by the model, and the fact that machines may be heterogeneous and
different from the ones used for the benchmarks.

% %%%%%%%%%%%%%%%%%%%%%%%
% \section{Experiments}
% \label{sec:expe}

%%%%%%%%%%%%%%%%%%%%%%%
%% \section{Related work}
%% \label{sec:related}

%%%%%%%%%%%%%%%%%%%%%%%
\section{Conclusion}
\label{sec:conclusion}

In this paper, we presented models for \galaxymaker's and \momaf's
execution time, and for \galaxymaker's output files's size.  Execution
time models give an estimation that is quite close to the real
execution time for input files having a reasonable size. For small
input files, the model returns an overestimate of the execution time,
but this is easily explained by cache mechanisms at the hardware
level, and can be taken into account in the model.

We also modeled the post-processing workflow, and provided efficient
means of executed it in a parameter sweep manner on a grid of
computers. Our solution relies on the \diet middleware that provides
transparent access to resources, and data and workflows
management. All the complexity of running the post-processing on a
grid, and the creation of the workflows are hidden by a web interface,
that provides an easy and user-friendly way of submitting such
cosmological simulations post-processing on many heterogeneous and
distributed resources.

The next step is of course to analyze the results produced by the
post-processing, and derive which parameters are the most influent on
the results. Finally, we aim at providing ranges of values for the
different parameters, that would provide correct results, \ie
comparable to observational data.

%%%%%%%%%%%%%%%%%%%%%%%%%%%%%%%%%%%%%%%%%%%%%%%%
%% BACKMATTER
%%%%%%%%%%%%%%%%%%%%%%%%%%%%%%%%%%%%%%%%%%%%%%%%

\begin{theacknowledgments}

This work was developed with financial support from the ANR (Agence
Nationale de la Recherche) through the LEGO project referenced
ANR-05-CIGC-11.

\end{theacknowledgments}


\bibliographystyle{aipproc}   % if natbib is available

\bibliography{paris}

\end{document}